\documentclass[pra,twocolumn,titlepage,nofootinbib,amsmath,showpacs
]{revtex4}%
\usepackage{amsmath}
\usepackage{amsfonts}
\usepackage{amssymb}
\usepackage{graphicx}%
\begin{document}
\title{\large A self-consistent value of the electric radius of the proton
from the Lamb shift in muonic hydrogen}
\author{Savely~G.~Karshenboim}
\email{savely.karshenboim@mpq.mpg.de}
\affiliation{Max-Planck-Institut f\"ur Quantenoptik, Garching,
85748, Germany} \affiliation{Pulkovo Observatory, St.Petersburg,
196140, Russia}



\begin{abstract}
Recently a high-precision measurement of the Lamb shift in muonic
hydrogen has been performed. An accurate value of the proton charge
radius can be extracted from this datum with a high accuracy. To do
that a sufficient accuracy should be achieved also on the theoretical
side, including an appropriate treatment of higher-order
proton-structure effects. Here we consider a higher-order contribution
of the finite size of the proton to the
Lamb shift in muonic hydrogen. Only model-dependent results for
this correction have been known up to date. Meantime, the involved
models are not consistent either with the existing experimental data
on the electron-proton scattering or with the value for the electric
charge radius of the proton extracted from the Lamb shift in muonic
hydrogen. We consider the higher-order contribution of the proton
finite size in a model-independent way and eventually derive a
self-consistent value of the electric radius of the proton.
The re-evaluated value of the proton charge radius is found to be
$R_E=0.840\,22(56)\;$fm.
\end{abstract}

\maketitle

\section{Introduction}

\begin{figure}[thbp]
\begin{center}
\resizebox{0.80\columnwidth}{!}{\includegraphics{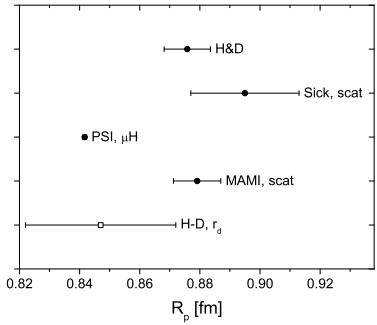}}
\end{center}
\caption{Determination of the rms proton charge radius.
For details see \cite{my_adp,my_ufn}.
}
\label{fig:re}       
\end{figure}

There is a controversy in a determination of the electric charge
radius of the proton. The most accurate value (as claimed) of the
proton charge radius comes from the measurements of the Lamb shift in
muonic hydrogen by the CREMA collaboration \cite{Nature,Science}. It
strongly disagrees with the scattering results \cite{mami,sick} (as
well as with the result from hydrogen and deuterium spectroscopy
summarized in \cite{codata2010}). The situation is summarized in
Fig.~\ref{fig:re} where two basic scattering results are presented.
Sick \cite{sick} has evaluated all the world data, but MAMI results
\cite{mami}. Other evaluations without MAMI data produced similar
results. MAMI results are presented in the plot separately.

Meantime, to extract the value of the proton radius  from the $\mu$H
Lamb shift one has to integrate the proton form factor over a broad
area in momentum space with a contribution from the low momenta
being enhanced.

A straightforward evaluation based on fitting of scattering data is
not appropriate since the results of such a fitting are not
consistent with the results on the muonic hydrogen Lamb shift.

Alternatively, one may use a certain model, such as the dipole
parametrization. Its parameter
has to be related to the proton radius. That makes the form factor
incompatible with the scattering data. (As it is well known, the
dipole parametrization is consistent with the higher-momentum
transfer data with the parameter inconsistent with any radius in
Fig.~\ref{fig:re}.)

Apparently, both these  options are not satisfactory. However, up to
date only such evaluations have been performed to extract the proton
radius from the muonic hydrogen (see, e.g., \cite{Nature,Science}).

The $\mu$H result for the proton radius was obtained for the first
time in \cite{Nature}. The theoretical expression for the Lamb shift
was presented there as
\begin{eqnarray}\label{th:nat}
E(2p_{1/2}-2s_{1/2}) &=& \bigl(
206.0573(45)_{\rm QED}
- 5.2262 r_p^2\nonumber\\
&& + 0.0347 r^3_p\bigr)\;{\rm meV}\;,
\end{eqnarray}
where $r_p=R_E/$fm is the numerical value of the proton charge
radius in the units of Fermi (=\,femtometer). The term denoted here
as `QED' is
dominated by the QED contributions, but contains also some small non-QED
terms such as the proton polarizability contribution.

Consider this expression in detail. In the model-independent terms,
one has to write rather
\begin{eqnarray}\label{th:nat:i3}
E(2p_{1/2}-2s_{1/2}) &=& \bigl(
206.0573(45)_{\rm QED}
- 5.2262 r_p^2
\bigr)\;{\rm meV}\nonumber\\
&& +\frac{2(Z\alpha)^5\,m_r^4}{\pi}\, I_3^{\rm E}\;,
\end{eqnarray}
where $m_r$ is the reduced mass. Here\footnote{The other notation
used is
\begin{eqnarray}
\langle r^3 \rangle_2 &=&\frac{48}{\pi}I_3^{\rm E}\nonumber\;.
\end{eqnarray}
The names are {\em the Friar momentum\/} or {\em the third Zemach
momentum\/}.}
\begin{eqnarray}
I_3^{\rm E}&\equiv&  \int_0^\infty
{\frac{dq}{q^4}}\left[\left(G_E(q^2)\right)^2-1-2G_E^\prime(0)\,q^2\right]\label{def:i3}\;,
\end{eqnarray}
is the integral we are interested in this paper. It describes the
next-to-leading
higher-order proton-finite-size contribution to the $\mu$H Lamb
shift. (The related contribution to the Lamb shift in ordinary
hydrogen is negligible.) We use the relativistic units in which
$\hbar=c=1$ throughout the paper.

The representation (\ref{th:nat}), applied in \cite{Nature}, was
obtained (see, e.g., \cite{EGS}) within the dipole parametrization
\[
G_{\rm dip}(q^2)=\left(\frac{\Lambda^2}{q^2+\Lambda^2}\right)^2\;,
\]
which describes the whole electric form factor with a single
parameter adjusted there to the value of the proton charge radius. As
it is well known, such a parametrization with
$\Lambda^2=0.71\,{\rm GeV}^2$ is a good one at higher momentum, but
it produces the radius which agrees neither with the
proton-scattering evaluation nor with the result from muonic
hydrogen. Meanwhile, while using a parameter consistent with the
muonic-hydrogen Lamb shift \cite{Nature}, the dipole parametrization
is not consistent with the experimental scattering data any more.
In other words, the low-momentum behavior of the form
factor, established through the measurement \cite{Nature}, and its
high-momentum behavior, established by scattering data, cannot be successfully
described by the dipole parametrization with a single parameter.

Actually, there is no parametrization of the proton form factor
which is literally correct and consistent with the data. Most of
the empiric parametrizations
\cite{kelly,as2007,am2007,ab2009,va2011} (see Appendix
\ref{s:fit} for detail) deal with the ratio of polynomials in $q^2$.
(Here, $q$ is the Euclidean momentum and negative values of $q^2$
represent the time-like region.) It is known that the form factor should
have a cut line at negative $q^2$ starting from $4m_\pi^2$. Meantime
a rational parametrization can produce only [a few] poles, but no
branch point. There is no uncertainty assigned to such a mismatch.
(There is a number of fits with even worse analytic behavior or
with wrong asymptotic behavior at high $q^2$.) Meantime, any
efforts to produce a theoretically motivated parametrization with a
correct position of the cut line and a correct discontinuity
function on the cut line (see, e.g.,\cite{lorenz}) are far from
good agreement with the experimental data (if we rely on the
$\chi^2$ criterion).

Below we look for a realistic estimation of the uncertainty in the
calculation of $I_3^{\rm E}$ and a
self-consistent determination of $R_E$ from muonic hydrogen.

To calculate the integral $I_3^{\rm E}$ in (\ref{def:i3}), we have
to integrate over the subtracted form factor,
\[
\left(G_E(q^2)\right)^2-1-2G_E^\prime(0)\,q^2\;.
\]
Obviously, we have
no direct experimental knowledge of it both at low and high momenta.
All that was used previously by various authors as the integrand was
a result of fitting rather than  direct measurements.

\begin{figure}[thbp]
\begin{center}
\resizebox{0.90\columnwidth}{!}{\includegraphics{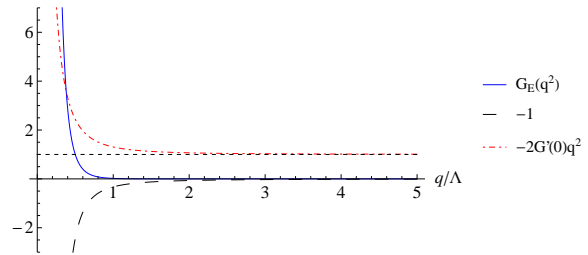}
}
\end{center}
\caption{Fractional contributions to the integrand in (\ref{def:i3})
as a function of $q/\Lambda$ as follows from the dipole model.
The red dot-dashed
line is for the subtraction term with $G_E^\prime(0)$. The dashed line is
for the subtraction term with 1 and the blue solid line is for the $G_E^2$
term, which should follow the experimental data. }
\label{fig:ind3}       
\end{figure}

The situation for an integration over experimental data is
illustrated in Fig.~\ref{fig:ind3}. We use there the dipole
parametrization to roughly estimate the scale of contributions of
separate terms. Various fractional contributions to the
integrand are presented as a function of $q/\Lambda$. The red
dot-dashed line is
for the subtraction term with $G^\prime_E(0)$, which is determined from a
fit and is to be directly related to $R_E^2$. The blue solid line is for
the $G_E^2$ term. This term should be determined by the experimental
data. The integral is fast
convergent with high $q$. Above $q=0.6 \Lambda$ the red dot-dashed line, which
is determined completely by the fit at zero momentum, is the
absolutely dominant contribution. Meanwhile at low $q$  separate
contributions are divergent and only their strong cancellation, which
can be successfully done only within a model, makes the integral
convergent.

We are going to split the integration into two parts:
\begin{equation}
I=\int_0^\infty {{dq}{...}}\equiv I_<+I_>\equiv
\int_0^{q_0}{{dq}{...}} + \int_{q_0}^\infty {{dq}{...}}
\end{equation}
which are to be treated differently. (We in part explore here an
idea suggested earlier in \cite{pla} for a somewhat similar
integral for the hyperfine splitting in hydrogen.)

At higher momenta, we will use `direct' experimental data. (We put
`direct' into the quotation marks because we rather intend to apply
in a certain way various appropriate empiric fits than the data
themselves.) The accuracy of the form factors is roughly 1\%. Once one
uses  the `direct data'  the
integral $I_>$ is indeed not really convergent at
$q_0\to0$, because the experimental value of $G_E(0)$ (or at any $q$
closely approaching to zero) is not equal to unity
--- it is only consistent with unity within the uncertainty. The
smaller is $q_0$, the larger is the uncertainty of the related
integral. It becomes divergent at $q_0\to 0$, unless one substitutes
the data by a fit, which we are not going to do in the low-momentum area.

Beside the integration over the data, there are also two subtraction
terms (see Fig.~\ref{fig:ind3}). One of them does not depend on the
fit and the data (it is related to the unity in the numerator in
(\ref{def:i3})) and the other, which is proportional to $G^\prime_E(0)$,
is presented in terms of the electric charge radius
\[
G^\prime_E(0)=-\frac16 R_E^2\;,
\]
which is not a parameter to be found from the fit of the scattering
data, but a constant to be determined from the eventual evaluation of
the Lamb shift in muonic hydrogen.

Thus we arrive at
\begin{equation}\label{q:data}
I_{3>}^{\rm E}\equiv I_{3>}^{\rm E}({\rm data})+\frac13 R_E^2\int_{q_0}^\infty\frac{dq}{q^2}\;.
\end{equation}

On the other hand, we can take advantage of expanding the form
factor at low momenta
\begin{equation}\label{g2}
\bigl(G_E(q^2)\bigr)^2= A + B q^2 + C q^4 + ...
\end{equation}
Some contributions into
\[
I_{3<}^{\rm E}\equiv  \int_0^{q_0}
{\frac{dq}{q^4}}\left[\left(G_E(q^2)\right)^2-1-2G_E^\prime(0)\,q^2\right]\label{def:i3<}\;,
\]
vanish because of the subtraction and
the uncertainty comes from the remaining terms. The smaller is $q_0$,
the smaller is the uncertainty. One finds $A=1$ and $B=-R_E^2/3$,
with the $A$ and $B$ terms canceling out. The leading non-vanishing
term is the $C$ term, which is now responsible for the contribution
and the uncertainty
of the integration over the low momenta.

The idea is to apply a certain model to estimate the uncertainties
and to find an optimal value of $q_0$ which corresponds to the
smallest uncertainty possible. Afterwards, we can apply a more
sophisticated description of the data and to find the related part
of $I_>$ by integrating over them.

As for the model to estimate the uncertainty, we note that the
dipole form factor provides a reasonable estimation for the form factor as
far as we discuss general features, but not any accurate particular
value. So, we can, e.g., set for (\ref{g2})
\[
C=C^{\rm dip}\times (1\pm1)\;,
\]
where $C^{\rm dip}$ is the dipole value.

The details of the estimation (whether, e.g., it should be $0\pm1$,
$1\pm0.5$, or $0.5\pm1$) may be discussed separately. $C$ is here
the curvature of the $(G_E(q^2))^2$ curve
and thus it is a certain general feature, which we expect, is
reasonably good presented by the dipole approximation. One may
expect that the dipole fit follows from the dispersion relations as
a simplified model for the dispersion density and thus reflects
physics (within certain margins).

\section{Consideration within the standard dipole model}

First, let us calculate the integral within the standard dipole
approximation, the result of which is indeed well known (see, e.g. \cite{EGS}):
\begin{eqnarray}\label{i3dip}
I_3^{\rm dip}&=&  \int_0^\infty {\frac{dq}{q^4}}\left[\left(G_{\rm dip}(q^2)\right)^2-1-2G_{\rm dip}^\prime(0)\,q^2\right]\nonumber\\
&=& \frac{105}{32}\frac\pi{\Lambda^3} \nonumber\\
&\simeq& 17.2\,{\rm GeV}^{-3}\nonumber\\
&\simeq&0.132\,{\rm fm}^3\;.
\end{eqnarray}

For numerical evaluations we use $\Lambda^2=0.71\;$GeV$^2/c^2$
($\Lambda=0.843\;$GeV). We also remind that
\[
R_{\rm dip}^2=\frac{12}{\Lambda^2}\;.
\]
The value related to $\Lambda^2=0.71\;$GeV$^2/c^2$ is 0.811\;fm,
which is indeed too low to be correct (cf. Fig.~\ref{fig:re}).

\section{Splitting the integral into parts}

The dipole fit is a good first approximation for a central value of
the form factor. The question is the accuracy. We assume that we
know the form factor for all areas of interest with accuracy at the
level of 1\%. That is indeed not sufficient to calculate the
integral (\ref{def:i3}) directly, but it can be applied for
estimation of the uncertainty of $I_>$ and eventually to find an
optimal value of the separation parameter $q_0$, which minimizes the total
uncertainty.

Let us start with $I_>$
\begin{eqnarray}
I_{3>}^{\rm E}&=&  \int_{q_0}^\infty {\frac{dq}{q^4}}\left[\left(G_E(q^2)\right)^2-1-2G_E^\prime(0)\,q^2\right]\nonumber\\
&=& \int_{q_0}^\infty
{\frac{dq}{q^4}}\left(G_E(q^2)\right)^2-\frac{1}{3q_0^3}-\frac{2G_E^\prime(0)}{q_0}\;.
\end{eqnarray}
The uncertainty comes from the first term only.
The subtractions do not contribute to the uncertainty as far as the
third term with $G_{\rm E}^\prime(0)$ is considered separately as in
(\ref{q:data}) (see below).

Thus to estimate the uncertainty we arrive at
\begin{eqnarray}
\delta I_{3>}^{\rm E}&=&  \delta\int_{q_0}^\infty {\frac{dq}{q^4}}\left(G_E(q^2)\right)^2\nonumber\\
&\simeq&  \delta\int_{q_0}^\infty {\frac{dq}{q^4}}\left(G_E(q_0^2)\right)^2\nonumber\\
&\simeq& \frac{1}{3(\nu \Lambda)^3}\frac{2\delta G_E(q_0^2)}{G_E(q_0^2)}\,\bigl(G_d(q_0^2)\bigr)^2 \nonumber\\
&=& \frac{1}{3(\nu \Lambda)^3}\frac{2\delta G_E(q_0)}{G_E(q_0^2)}
\left(\frac{1}{1+\nu^2}\right)^4\;.
\end{eqnarray}
where
\[
\nu=\frac{q_0}{\Lambda}\;.
\]
Here, we suggest that the uncertainty comes only from integration
around the lower limit, where the form factor can be roughly
estimated by the dipole fit.

To better understand a possible outcome qualitatively, it is useful to consider
rather relative contributions  than the absolute ones. Since the exact value
used for the normalization does not play any real role here, we
apply the dipole value (\ref{i3dip}) of $I_3$ for this purpose. In
particular, assuming that we experimentally know the form factor
within 1\% uncertainty, we find
\begin{equation}\label{unc>}
\frac{\delta I_{3>}^{\rm E}}{I_3^{\rm E}}\simeq \frac{0.000\,65}{
\nu^3}\left(\frac{1}{1+\nu^2}\right)^4\;.
\end{equation}

As we mention above, the $G^\prime(0)$ contribution needs a specific
treatment. That contribution
\begin{eqnarray}
I_3^{\rm R}&\equiv&  \int_{q_0}^\infty
{\frac{dq}{q^4}}\left[-2G_E^\prime(0)q^2\right]\nonumber\\
&=&\frac13\int_{q_0}^\infty {\frac{dq}{q^2}}R_E^2\nonumber\\
&=&\frac13\frac{R_E^2}{q_0}
\end{eqnarray}
should `renormalize' the $r^2_p$ coefficient in the theoretical
expression, which now reads as
\begin{eqnarray}
E(2p_{1/2}-2s_{1/2}) &=& \bigl(209.9779(45)_{\rm QED} - 5.2226 r_p^2
\bigr)\;{\rm meV}\nonumber\\
&& +\frac{2(Z\alpha)^5m_r^4}{3\pi}\,\frac{R_E^2}{q_0}\nonumber\\&&
+\frac{2(Z\alpha)^5m_r^4}{\pi}\, \bigl(I_3^{\rm E}-I_3^{\rm
R}\bigr)\;.
\end{eqnarray}

Now, let us consider $I_{<}$. At low $q^2$ we find for $G_{\rm
dip}^2$
\[
\left(\frac{\Lambda^2}{\Lambda^2+q^2}\right)^4
=1-4\frac{q^2}{\Lambda^2}+10 \left(\frac{q^2}{\Lambda^2}\right)^2-20
\left(\frac{q^2}{\Lambda^2}\right)^3+\dots
\]
and we suggest for the `real' form factor
\begin{eqnarray}
\left(G_E(q_0^2)\right)^2 &=&1-4a\frac{q^2}{\Lambda^2}+10b
\left(\frac{q^2}{\Lambda^2}\right)^2
\end{eqnarray}
the coefficients $a$ and $b$ are not too far from the unity. We are to set here
$b=1\pm1$. We denote $\pm 1$ as $\pm\delta b$.

So, we find
\begin{eqnarray}
I_{3<}^{\rm E}&\simeq&  10b \,\int_0^{q_0}{\frac{dq}{\Lambda^4}}\nonumber\\
&=& 10\,(1\pm\delta b) \, \frac{\nu}{ \Lambda^3}
\end{eqnarray}
and
\begin{equation}\label{unc<}
\frac{\delta I_{3<}^{\rm E}}{I_3^{\rm E}}\simeq 0.97 \,\delta b \,\nu\;.
\end{equation}

\begin{figure}[thbp]
\begin{center}
\resizebox{0.85\columnwidth}{!}{\includegraphics{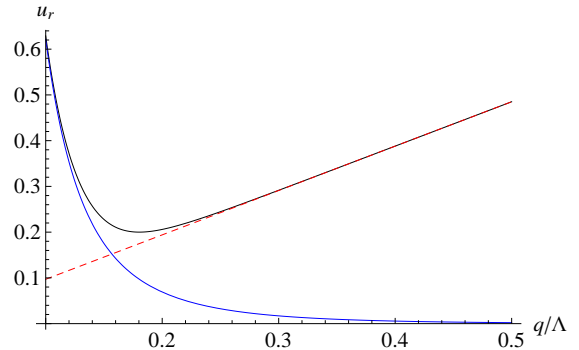}}
\end{center}
\caption{The final relative uncertainty as the rms sum of
contributions of (\ref{unc>}) and (\ref{unc<}) plotted as a function
of $\nu=q_0/\Lambda$. The partial uncertainties (\ref{unc>}) and
(\ref{unc<}) for $I_>$ and $I_<$ are also presented. The red dashed
line is for $\delta I_<$ and the blue solid line is for $\delta I_>$.
The relative uncertainties are estimated assuming that the central
value is determined by the standard dipole fit.}
\label{fig:unc3}       
\end{figure}

Finally, we obtain
\[
\frac{\delta I_{3}^{\rm E}}{I_3^{\rm E}}\simeq20\%
\]
at $q_0\simeq 0.1803\,\Lambda=0.152\,{\rm GeV}/c$ which is
(approximately) the best choice (see Fig.~\ref{fig:unc3} and
Table~\ref{t:un3d}).

\begin{table}[htbp]
\begin{center}
\begin{tabular}{l|c|c|c}
\hline
contribution & $\delta I_{3<}^{\rm E}/I_3^{\rm dip}$   & $\delta (I_{3>}^{\rm E}-I_3^{\rm R})/I_3^{\rm dip}$ & total  \\[0.8ex]
\hline
relative uncertainty  & 17.5\% &  9.7\%  & 20.0\%  \\[0.8ex]
\hline
\end{tabular}
\caption{Relative uncertainty budget for $I_3^{\rm E}-I_3^{\rm R}$
at $q_0\simeq 0.1803\,\Lambda=0.152\,{\rm GeV}/c$ from the dipole
consideration.\label{t:un3d}}
 \end{center}
 \end{table}

The $q_0$ dependence is rather flat and in the interval at
$\nu_0=0.15-0.3$ (that relates to $q_0=(0.12-0.25)\;$GeV/$c$) the
value of ${\delta I_{3}^{\rm E}}/{I_3^{\rm E}}$ is not bigger than 25\% (see
Table~\ref{t:un3d:q}).

\begin{table}[htbp]
\begin{center}
\begin{tabular}{l|c|c|c|c|c}
\hline
 $q_0/\Lambda$ &$q_0$ & $\delta I_{3<}^{\rm E}/I_3^{\rm dip}$ & $\delta (I_{3>}^{\rm E}-I_3^{\rm R})/I_3^{\rm dip}$ & total &{\em scatter} \\[0.8ex]
\hline
 0.10 & 0.084\;GeV & 9.7\% &  62\%  & 63\% & {\em 11\%} \\[0.8ex]
 0.15 & 0.126\;GeV & 14.6\% &  17.5\%  & 22.3\% & {\em 5.5\%} \\[0.8ex]
 0.20 & 0.169\;GeV &  19.4\% & 6.9\%   & 20.6\% & {\em 3.0\%} \\[0.8ex]
 0.25 & 0.211\;GeV &  24.2\% & 3.2\%   & 24.4\% & {\em 1.7\%} \\[0.8ex]
 0.30 & 0.253\;GeV &  29.1\% & 1.7\%   & 29.2\% & {\em 0.9\%} \\[0.8ex]
 0.40 & 0.337\;GeV &  38.8\% & 0.6\%   & 38.8\% & {\em 0.3\%} \\[0.8ex]
\hline
\end{tabular}
\caption{Relative uncertainty budget for $I_3^{\rm E}-I_3^{\rm R}$
at various $q_0/\Lambda$  from the dipole consideration.  The last
column is for the scatter in units of $I^{\rm dip}$
(see Sect.~\ref{s:int_fit}) It is shown in Italic because it is not
included into the error budget, but used to control the uncertainty.
\label{t:un3d:q}}
 \end{center}
 \end{table}

The uncertainty in $I_3^{\rm }$ at the level of 20\% should affect
the uncertainty of $R_E$ from the Lamb shift in muonic hydrogen, but
not dramatically.

\section{Integration over the fits\label{s:int_fit}}

It would be indeed preferable to calculate $I_{3>}$ with real
scattering data, which is unfortunately not that easy. Here, we use
another opportunity and apply fits. However, we should distinguish a
fit in an area, where we really have data, and a fit outside of such an
area. The latter concerns not only the kinematic area, but also the
accuracy. Indeed, we have certain data at low momentum transfer, but
we have no direct data with sufficient accuracy for the subtracted
form factor to calculate $I_3$.

We intend to work with the fits in an area where all model-dependent
effects are negligible. A reasonable estimation of the systematic
uncertainty due to choice of the fits can be estimated by utilizing
fits with consistent
behavior at high and medium momentum transfer, but with different
behavior at low momenta.

As an approximation, we apply fits for the electric form factor of
the proton from Kelly, 2004 \cite{kelly},
Arrington and Sick, 2007 \cite{as2007}, Arrington et
al., 2007 \cite{am2007}, Alberico et al., 2009 \cite{ab2009}, Venkat
et al., 2011 \cite{va2011}, and Bosted, 1995 \cite{bo1994}.

Two of them are with so-called chain fractions, four fits are with
Pad\'e approximations with polynomials in $q^2$ and one is a Pad\'e
approximation with polynomials in $q$ (see Appendix \ref{s:fit}).

\begin{figure}[thbp]
\begin{center}
\resizebox{0.85\columnwidth}{!}{\includegraphics{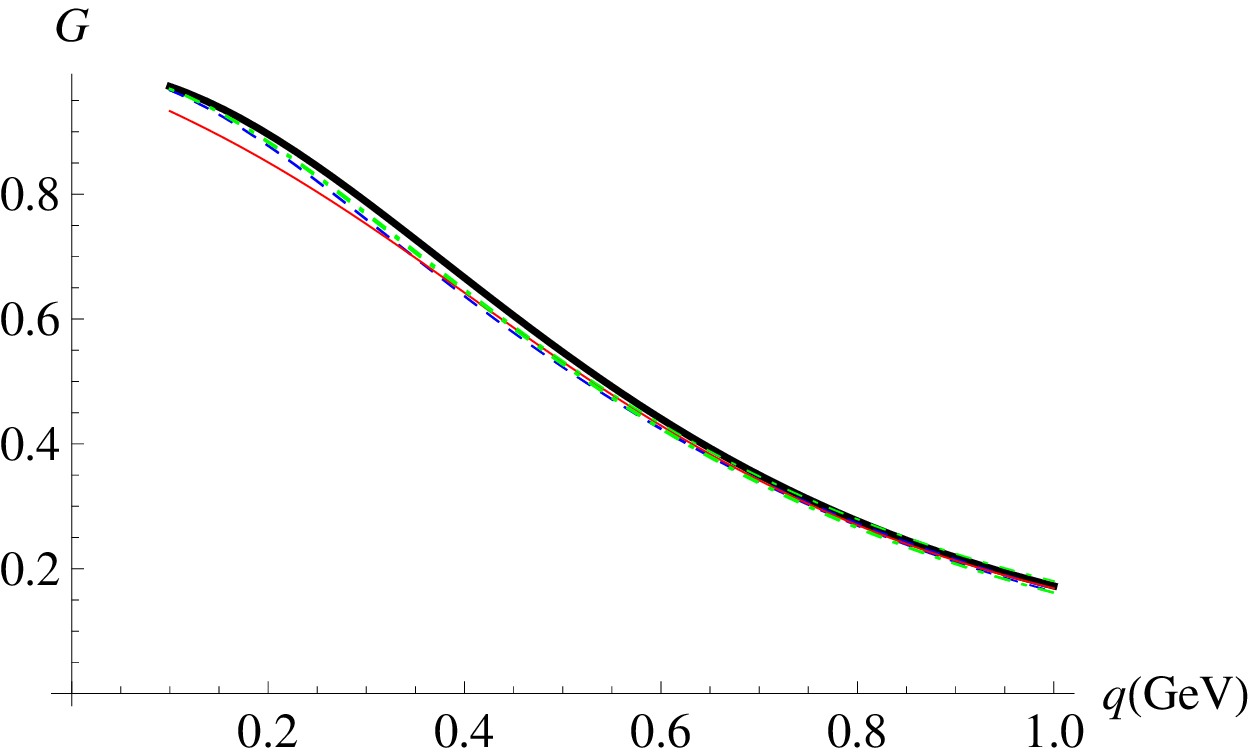}}
\resizebox{0.85\columnwidth}{!}{\includegraphics{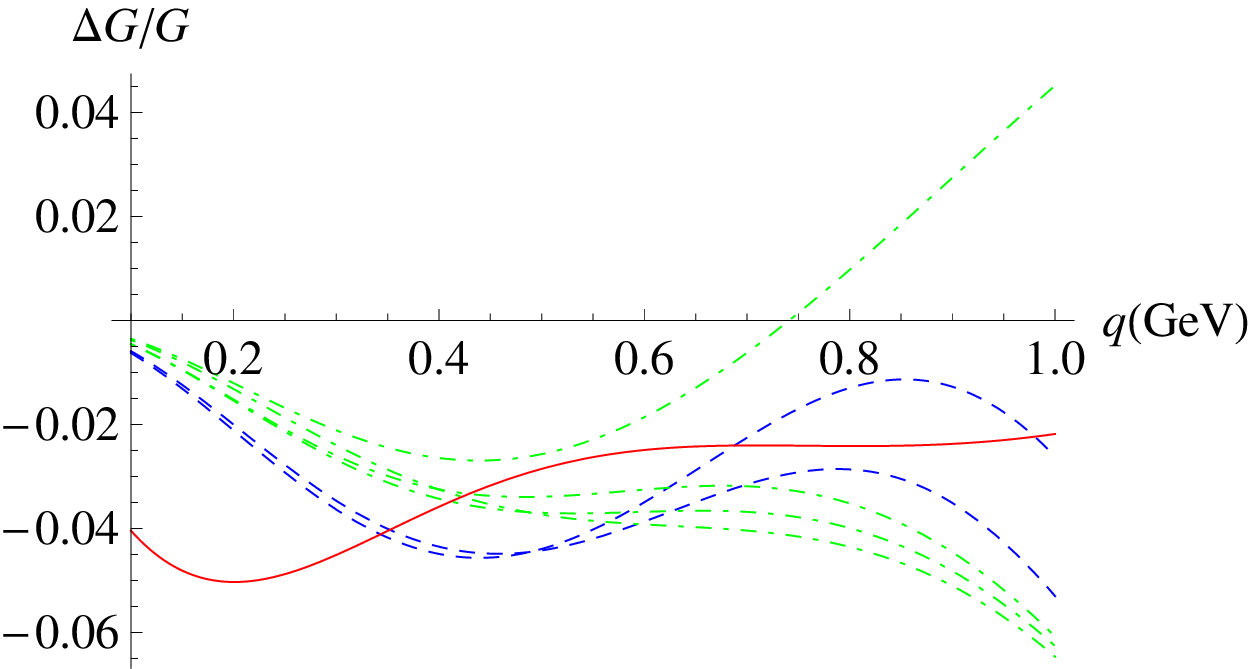} }
\end{center}
\caption{The electric-charge form factor of the proton $G_E(q^2)$.
Top: the dipole parametrization and the fits from
\cite{kelly,as2007,am2007,ab2009,va2011,bo1994} (see
Appendix~\ref{s:fit} for details). Bottom: Relative deviation of
the fits from the dipole form factor, $(G_E-G_{\rm dip})/G_{\rm
dip}$. Horizontal axis: $q\;$[Gev/$c$]. Blue dashed lines are for the chain
fractions, green dot-dashed lines are for Pad\'e approximations with
$\tau=q^2/4m_p^2$ and the red solid one is for the Pad\'e approximation
with $q$. The standard dipole approximation is  presented with a black
 bold solid line in the top graph.}
\label{fig:edip}       
\end{figure}

The fits are quite close one to another and to the standard
dipole parametrization in area of interest. Their comparison is
presented in Fig.~\ref{fig:edip}.

The low-momentum behavior of the fits is summarized in
Table~\ref{t:exp:fits}. We see that the Pad\'e fits tend to have a
somewhat lower value of the radius and of the $C$ coefficient, than
those for the chain-fraction fits. All the radius values are
substantially above the one from muonic hydrogen (see
Fig.~\ref{fig:re}). The coefficient for the $q^4$ term for
$\bigl(G_E(q^2)\bigr)^2$ is quite above that for the standard dipole
parametrization, but within the margins ($1\pm1$), we have applied
in our evaluation.

\begin{table}[htbp]
\begin{center}
\begin{tabular}{l|c|c|c|c}
\hline
~fit & ref. & type &  $R_E$ [fm] &  ~$C$ [GeV$^{-4}$]\\[0.8ex]
\hline
(\ref{fit:as2007}) & \cite{as2007}& chain fraction & 0.90 & 34.3 \\[0.8ex]
(\ref{fit:bm2005}) & \cite{as2007}& chain fraction & 0.90 & 35.3 \\[0.8ex]
(\ref{fit:kelly}) & \cite{kelly}& Pad\'e approximation ($q^2$) & 0.86 & 28.0 \\[0.8ex]
(\ref{fit:am2007}) & \cite{am2007}& Pad\'e approximation ($q^2$) & 0.88 & 31.1 \\[0.8ex]
(\ref{fit:ab2009}) & \cite{ab2009}& Pad\'e approximation ($q^2$) & 0.87 & 28.2 \\[0.8ex]
(\ref{fit:va2011}) & \cite{va2011}& Pad\'e approximation ($q^2$) & 0.88 & 31.3 \\[0.8ex]
\hline
\end{tabular}
\caption{The low-momentum expansion of the fits for the electric form
factor of the proton. The values
are given for central values of the fits without any uncertainty. Here:
$\bigl(G_E(q^2)\bigr)^2= 1 - R_E^2 q^2/3 + C q^4 + ...$.
The related values for the standard dipole fit are $R_E=0.811\;$fm
and $C=19.8\;{\rm GeV}^{-4}$. The spread of the central values of the
charge radius (from 0.86 to 0.90\;fm) is comparable with spread of
central values of the results in Fig~\ref{fig:re}, which are
0.84 ($\mu$H), 0.88 (H{\&}D), 0.895 (Sick) and 0.88 (MAMI) fm.
\label{t:exp:fits}}
 \end{center}
 \end{table}

One of the fits (from \cite{bo1994}) is not included into
Table~\ref{t:exp:fits}. That is a Pad\'e approximation with
polynomials in $q$ (see Eq. (\ref{fit:bo1994})). Containing terms linear
in $q$  in the denominator, the fit definitely has a low-momentum
behavior, strongly different from all the others. (Nothing is
incorrect with the fit. It was designed as a phenomenological fit
for not very low momentum transfer and is consistent there with the
data.) Such an `inappropriate' behavior of the fit makes it to be a
perfect tool for tests on the model dependence related to the
assumption on the low-momentum behavior.

The result of integration
\[
I_{3>}^{\rm E} - I_{3}^{\rm R}=\int_{q_0}^\infty
{\frac{dq}{q^4}}\left[\left(G(q^2)\right)^2-1\right]
\]
over the different fits with the cut-off parameter $q_0\simeq
0.1803\,\Lambda=0.152\,{\rm GeV}/c$ lays  from $-25.8\;{\rm
GeV}^{-3}$ to $-24.5\;{\rm GeV}^{-3}$ (if we exclude integration
over the fit with the Pad\'e approximation in $q$) and from $-29.6\;{\rm
GeV}^{-3}$ (if we include it). For details, see
Table~\ref{t:i3:fits}.

\begin{table}[htbp]
\begin{center}
\begin{tabular}{l|c|c|c}
\hline
~fit & ref. & type &  $I_{3>}^{\rm E}-I_3^{\rm R}$  \\[0.8ex]
\hline
(\ref{fit:as2007}) & \cite{as2007}& chain fraction & $-25.7\;{\rm GeV}^{-3}$\\[0.8ex]
(\ref{fit:bm2005}) & \cite{as2007}& chain fraction & $-25.8\;{\rm GeV}^{-3}$\\[0.8ex]
(\ref{fit:kelly}) & \cite{kelly}& Pad\'e approximation ($q^2$) & $-24.5\;{\rm GeV}^{-3}$\\[0.8ex]
(\ref{fit:am2007}) & \cite{am2007}& Pad\'e approximation ($q^2$) & $-25.0\;{\rm GeV}^{-3}$\\[0.8ex]
(\ref{fit:ab2009}) & \cite{ab2009}& Pad\'e approximation ($q^2$) & $-24.7\;{\rm GeV}^{-3}$\\[0.8ex]
(\ref{fit:va2011}) & \cite{va2011}& Pad\'e approximation ($q^2$) & $-25.0\;{\rm GeV}^{-3}$\\[0.8ex]
(\ref{fit:bo1994}) & \cite{bo1994}& Pad\'e approximation ($q$) & $-29.6\;{\rm GeV}^{-3}$\\[0.8ex]
\hline
\end{tabular}
\caption{Scatter of the results of numerical integration over the fits for $I_{3>}^{\rm E}-I_3^{\rm R}$ at
`optimal' $q_0\simeq
0.1803\,\Lambda=0.152\,{\rm GeV}/c$.\label{t:i3:fits}}
 \end{center}
 \end{table}

We consider
\[
I_{3>}^{\rm E} - I_{3}^{\rm R}=- 25.2(6)\;{\rm GeV}^{-3}
\]
as the estimation. The scatter of $I_3^{\rm E}-I_{3}^{\rm R}$ ($\pm
3.8\%$ if we exclude the fit with the Pad\'e approximation in $q$) is
below the total projected uncertainty and below the uncertainty for
$I_{3>}^{\rm E}-I_3^{\rm R}$. Once we include the fit with the Pad\'e
approximation in $q$, the scatter becomes 14.8\% which is
comparable with the total uncertainty. Here, we show the scatter in
various Tables, but do not include it into the error budget (see
Table~\ref{t:un3f}). We denote the scatter without
(\ref{fit:bo1994}) as {\em scatter\/} and the scatter with
(\ref{fit:bo1994}) as {\em scatter\/}$^*$. We define the `scatter'
in the Tables as a half value of the difference between the maximal
and minimal values. More discussion can be found in the conclusion section.

\begin{table}[htbp]
\begin{center}
\begin{tabular}{l|c|c|c|c}
\hline
contribution          & $\delta I_{3<}^{\rm E}$          & $\delta (I_{3>}^{\rm E}-I_3^{\rm R})$  & scatter& scatter$^*$  \\[0.8ex]
\hline
dipole consideration   & 17.5\% &  9.7\%   &  &  \\[0.8ex]
integration over fits &  &   &  3.8\% & 14.8\%\\[0.8ex]
\hline
\end{tabular}
\caption{The uncertainty and scatter for $I_3^{\rm E}-I_3^{\rm R}$
at $q_0\simeq 0.1803\,\Lambda=0.152\,{\rm GeV}/c$. \label{t:un3f}}
 \end{center}
 \end{table}

We mentioned above that the fit (\ref{fit:bo1994}) from \cite{bo1994}
has no reasonable behavior at low momentum. Nevertheless, its use
leads to a quite reasonable result which means that we are in a safe
area of a model-independent application of the fits.

We remind that we have considered the $I_{3<}^{\rm E}(\nu)$ in the
previous section and the result for $\nu_0=0.1803$ gives
\begin{eqnarray}
I_{3<} (\nu_0)
&=&
10\,(1\pm\delta b) \, \frac{\nu_0}{ \Lambda^3}\nonumber\\
&=& 3.0(3.0)\;{\rm GeV}^{-3}\;.
\end{eqnarray}

Finally we obtain
\[
I_{3}^{\rm E} - I_{3}^{\rm R}= -22.2(3.4)\;{\rm GeV}^{-3}\;.
\]

The integral can be calculated for various $q_0$ (see, e.g., the
results in Table~\ref{t:un3f:q}). The result for $I_{3}^{\rm E} -
I_{3}^{\rm R}$ depends on $q_0$, because the subtracted contribution
($I_{3}^{\rm R}$) explicitly depends on $q_0$. It is interesting to
compare the scatter of $I_{3>}^{\rm E}$ and the uncertainty. If we
go to smaller values of $q_0$, the variation of $G_E(q_0)$ from one
fit to another
becomes bigger than the uncertainty. For higher $q_0$ the
scatter reduces and the fit (\ref{fit:bo1994}) with the Pad\'e
approximation in $q$ \cite{bo1994} becomes more reasonable. The
reduction of the scatter is due to the fact that the integrand for
$I_>$ is proportional $G_E^2-1$ and the contribution of the fit (i.e. of
the $G_E^2$ term) becomes less and less important in comparison with
the unity (see, e.g., Fig.~\ref{fig:ind3}), while the integral
converges fast.

\begin{table}[htbp]
\begin{center}
\begin{tabular}{l|c|c|c}
\hline
$q_0/\Lambda$ & $I_{3}^{\rm E}-I_3^{\rm R}$ & ~~scatter of $I_{3>}^{\rm E}$& ~~scatter$^*$ of $I_{3>}^{\rm E}$\\[0.8ex]
\hline
0.10 &   $-57(11)$\;GeV$^{-3}$ & 2.0\;GeV$^{-3}$ & 18\;GeV$^{-3}$  \\[0.8ex]
0.15 &   $-30.8(3.9)$\;GeV$^{-3}$ & 1.0\;GeV$^{-3}$ & 5.0\;GeV$^{-3}$  \\[0.8ex]
0.20 &   $-18.0(3.5)$\;GeV$^{-3}$ & 0.5\;GeV$^{-3}$ & 1.7\;GeV$^{-3}$  \\[0.8ex]
0.25 &   $-10.4(4.2)$\;GeV$^{-3}$ & 0.3\;GeV$^{-3}$ & 0.6\;GeV$^{-3}$  \\[0.8ex]
0.30 &   $-5.4 (5.0)$\;GeV$^{-3}$  & 0.2\;GeV$^{-3}$ & 0.2\;GeV$^{-3}$  \\[0.8ex]
0.40 &   $+0.9 (6.7)$\;GeV$^{-3}$  & 0.06\;GeV$^{-3}$ & 0.06\;GeV$^{-3}$  \\[0.8ex]
\hline
\end{tabular}
\caption{Results for of $I_{3}^{\rm E}-I_3^{\rm R}$ and its scatter
from fit to fit at various values of $q_0$.\label{t:un3f:q}}
 \end{center}
 \end{table}

\section{The extraction of the $R_E$ value from the Lamb shift}

The purpose of this paper is not to obtain the proton charge radius
from an {\em ab initio\/} analysis of the Lamb shift in muonic
hydrogen, but to
calculate the {\em shift\/} in the value of the proton radius due to
a self-consistent treatment of the leading higher-order
proton-finite-size contribution\footnote{Indeed, the $I_3$
contribution is only the leading proton-finite-size contribution
beyond the $R_E^2$ term. There are additional smaller contributions,
such as a recoil correction to the higher-order $I_3$ term, the
proton-polarizability contribution etc. (see, e.g., \cite{carlson}).
They are in part included
into the `QED' term in (\ref{th:nat:i3}).}. The shift is in respect
 to already existing evaluations. Here, we first derive an
expression for the shift of $R_E$ and next discuss existing
extractions.

We remind that the leading finite-nuclear-size contribution to the
$ns$ energy is
\begin{equation}\label{eq:lead}
\frac{2}{3}\frac{(Z\alpha)^4}{n^3}m_r^3 R_E^2
\end{equation}
while the higher-order correction is
\begin{equation}
-\frac{16}{\pi}\frac{(Z\alpha)^5}{n^3}m_r^4 I_3^{\rm E}\;.
\end{equation}
For $l\neq0$ both contributions are zero. For the application to
measured transitions in muonic hydrogen \cite{Nature,Science},
$n=2$.

Eq. (\ref{eq:lead}) is not a complete result for the $R_E^2$ term,
used in (\ref{th:nat}) and (\ref{th:nat:i3}), because of an
$\alpha$-correction to the leading term (see, e.g., \cite{EGS}). We
deliberately ignore it here. The shift we are interested in, is not
much affected by such a correction to the leading term and ignoring
the correction we are well within uncertainty of our treatment for
the shift.

After applying the self-consistent approach developed above,
both finite-nuclear-size terms are effectively `re-normalized'
\begin{eqnarray}
\frac{2}{3}\frac{(Z\alpha)^4}{n^3}m_r^3 R_E^2 &\to&
\frac{2}{3}\frac{(Z\alpha)^4}{n^3}m_r^3\left[1-\frac{8(Z\alpha)}{\pi}\frac{m_r}{q_0}\right] R_E^2\;,\nonumber\\
-\frac{16}{\pi}\frac{(Z\alpha)^5}{n^3}m_r^4 I_3^{\rm E} &\to&
-\frac{16}{\pi}\frac{(Z\alpha)^5}{n^3}m_r^4 \bigl(I_3^{\rm
E}-I_3^{\rm R}\bigr)\;,
\end{eqnarray}

Denoting the `original' values of $R_E$ and $I_3^{\rm E}$ (from the
existing evaluations) as $R_E^0$ and $I_3^0$, we find
\begin{eqnarray}
&&\frac{2}{3}\frac{(Z\alpha)^4}{n^3}m_r^3 (R_E^0)^2-\frac{16}{\pi}\frac{(Z\alpha)^5}{n^3}m_r^4 I_3^{0}\nonumber\\
&=&\frac{2}{3}\frac{(Z\alpha)^4}{n^3}m_r^3\left[1-\frac{8(Z\alpha)}{\pi}\frac{m_r}{q_0}\right](R_E)^2\nonumber\\
&&-\frac{16}{\pi}\frac{(Z\alpha)^5}{n^3}m_r^4 \bigl(I_3^{\rm
E}-I_3^{\rm R}\bigr)\;.
\end{eqnarray}
The combination in the left-hand-side of the identity is determined
by a comparison of the experimental
value with the `QED' contribution in (\ref{th:nat}). Eventually we
obtain
\begin{eqnarray}\label{eq:extract3}
&&(R_E)^2-(R_E^0)^2 =\frac{1}{\left[1-\frac{8(Z\alpha)}{\pi}\frac{m_r}{q_0}\right]}\left[\frac{8(Z\alpha)}{\pi}\frac{m_r}{q_0}(R_E^0)^2\right.\nonumber\\
&+&\left.\frac{24(Z\alpha)m_r I_3^{0}}{\pi} \frac{I_3^{\rm
E}-I_3^{\rm R}-I_3^{0}}{I_3^{0}}\right]\;.
\end{eqnarray}

To determine the shift, we have to consider, how $I_3^{0}$ was
evaluated. We discuss below two CREMA's publications\footnote{We do
not examine the extractions by themselves there. We consider only a
shift due to change of the approach in treatment of $I_3^{\rm E}$.
Indeed, we understand that the results in Refs. \cite{Nature,Science}
are in part not compatible, because a more recent publication includes
certain updates of theory.}
\cite{Nature,Science}, where the most important extractions of the
proton charge radius were made and two different treatments of the
integral under question have been applied. The results of our
re-evaluation are summarized in Table~\ref{t:crema}.

\begin{table}[htbp]
\begin{center}
\begin{tabular}{l|c|c|c|c}
\hline
ref. & $R_E^0$ [fm] & $I_3^{0}$ [GeV$^{-3}$]& $R_E-R_E^0$  [fm]& $R_E$  [fm]\\[0.8ex]
\hline
\cite{Nature} & 0.841\,84(67) & 19.25 & $-0.000\,19(43)$ & 0.841\,65(79) \\[0.8ex]
\cite{Science} & 0.840\,87(39) & 22.9(1.2) & $-0.000\,65(43)(15)$ & 0.840\,22(56) \\[0.8ex]
\hline
\end{tabular}
\caption{Original parameters from CREMA's extractions in
\cite{Nature} and \cite{Science} and the corrections to the charge
radius due to the self-consistent evaluation of $I_3^{\rm E}$. The
uncertainties for the shift for the re-evaluation of the result from
\cite{Science} are explained in the text.\label{t:crema}}
 \end{center}
 \end{table}

Let us consider the previously applied evaluations in more detail.
In case of the first extraction \cite{Nature} of the proton radius
from muonic hydrogen Lamb shift, the model for a calculation of the
higher-order term suggested to apply a dipole shape of the form
factor, but with an adjustable parameter. The parameter was linked to
the radius. For more detail see, e.g., \cite{EGS}.

Technically, the result was
\[
I_3^{0}=\frac{105\pi}{32} \left(\frac{(R_E^0)^2}{12}\right)^{3/2}\;.
\]

The result for the most optimal value ($q_0\simeq
0.1803\,\Lambda=0.152\,{\rm GeV}/c$) is
\begin{eqnarray}\label{eq:newre}
R_E &=& 0.841\,65(79)\;{\rm fm} \nonumber\\
R_E-R_E^{\rm old} &=& -0.000\,19(43)\;{\rm fm}\;,
\end{eqnarray}
and the scatter due to choice of the fit is $0.000\,08\;{\rm fm}$.
The extraction for various $q_0$ is presented in
Table~\ref{t:extract3:q}.

\begin{table}[htbp]
\begin{center}
\begin{tabular}{l|c|c|c}
\hline
$q_0/\Lambda$ & $R_E$ &$R_E-R_E^{\rm old}$& ~~scatter\\[0.8ex]
\hline
0.10 &   0.841\,34(150)\;fm & $-0.000\,49(135)$\;fm& 0.000\,24\;fm \\[0.8ex]
0.15 &   0.841\,59(83)\;fm & $-0.000\,25(49)$\;fm& 0.000\,12\;fm \\[0.8ex]
0.20 &   0.841\,68(80)\;fm & $-0.000\,16(44)$\;fm & 0.000\,06\;fm\\[0.8ex]
0.25 &   0.841\,73(85)\;fm & $-0.000\,11(52)$\;fm & 0.000\,04\;fm\\[0.8ex]
0.30 &   0.841\,76(91)\;fm & $-0.000\,08(62)$\;fm & 0.000\,02\;fm\\[0.8ex]
0.40 &   0.841\,80(106)\;fm & $-0.000\,04(82)$\;fm & 0.000\,007\;fm\\[0.8ex]
\hline
\end{tabular}
\caption{The results for $R_E$ at various values of $q_0$ as follows
from (\ref{eq:extract3}). The scatter is related to the scatter of
$I_{3>}^{\rm E}$ in Table~\ref{t:un3f:q}.\label{t:extract3:q}}
 \end{center}
 \end{table}

The results obtained at various $q_0$ are in a perfect agreement
with each other and with (\ref{eq:newre}) and all have comparable
uncertainty, which is a strong confirmation of consistency of our
method.

Above we have found a new value of the proton radius re-evaluating
data from \cite{Nature}. That is not the best one. A better and
somewhat more reliable result was published by the CREMA collaboration
later in \cite{Science} on base of evaluation of their data from two
$2s-2p$ transitions in muonic hydrogen. The QED theory was updated
there as well as an evaluation of certain higher-order proton-structure
effects was included. The original result of \cite{Science} is
\[
R_E = 0.840\,87(39)\;{\rm fm}\;.
\]
The uncertainty of $0.000\,31\;$fm comes from the experiment and
$0.000\,29\;$fm is from theory, which, in particular, includes
the uncertainties of the proton-polarizability and elastic-two-photon
contributions (see Appendix \ref{s:sci} for detail). The
elastic two-photon contribution includes various recoil corrections,
but the dominant part is still related to $I_{3}^{\rm E}$, which
involves some uncertainty due to a model-dependent evaluation of
$I_3^{\rm E}$. We have to find a new central value as explained above and to
re-evaluate the uncertainty.

The results of re-evaluation of both CREMA's results are summarized
in Table~\ref{t:crema}. The value of $R_E-R_E^0$ presented there
has an uncertainty due to our re-evaluation of $I_3^{\rm E}$, which
is 0.000\,43\;fm. It is the same for the re-evaluation of both CREMA's
results. However, the later result has an uncertainty due to the
evaluation of $I_3^0$, which is 0.000\,15\;fm. To check the
consistency of the evaluation of $I_3^{\rm E}$ \cite{carlson} used in
\cite{Science} with ours, we have to combine the uncertainties, That
leads to the overall uncertainty of 0.000\,45\;fm. Meantime, to
correct the radius, we have to remove the uncertainty of
0.000\,15\;fm of the former evaluation and to include 0.000\,43\;fm
of ours.

Similarly to the consideration of the re-evaluation of the result
from \cite{Nature}, we have calculated the result for the corrected
radius for \cite{Science} at different values of the separation
parameter $q_0$. The results are summarized in Table~\ref{t:science}.
The results for the shift $R_E-R_E^{\rm old}$ at various values
of $q_0$ are consistent.

\begin{table}[htbp]
\begin{center}
\begin{tabular}{l|c|c|c}
\hline
$q_0/\Lambda$ & $R_E$ &$R_E-R_E^{\rm old}$& ~~scatter\\[0.8ex]
\hline
0.10 &   0.839\,90(140)\;fm & $-0.000\,98(135)(15)$\;fm& 0.000\,24\;fm \\[0.8ex]
0.15 &  0.840\,15(62)\;fm &  $-0.000\,73(49)(15)$\;fm & 0.000\,12\;fm\\[0.8ex]
0.20 &  0.840\,25(59)\;fm &  $-0.000\,62(44)(15)$\;fm & 0.000\,06\;fm\\[0.8ex]
0.25 & 0.840\,30(65)\;fm &  $-0.000\,58(52)(15)$\;fm & 0.000\,04\;fm\\[0.8ex]
0.30 & 0.840\,33(73)\;fm &  $-0.000\,54(62)(15)$\;fm & 0.000\,02\;fm\\[0.8ex]
0.40 & 0.840\,37(91)\;fm &  $-0.000\,50(82)(15)$\;fm & 0.000\,007\;fm\\[0.8ex]
\hline
\end{tabular}
\caption{The results for the correction to the charge radius of the
  proton
at various values of $q_0$ for the extraction from \cite{Science}
(cf. Table~\ref{t:extract3:q}).\label{t:science}}
\end{center}
\end{table}

\section{Comparison with former extractions}

Comparing the corrections we conclude that the model applied to
the calculation of $I_3$ in the earlier CREMA publication \cite{Nature}
happened to be somewhat more successful that the one applied later
in \cite{Science}. Meantime, the latter model is more realistic.
Let us briefly examine the correction in both cases and next explain why one
description is somewhat more successful than the other.

The shift for the result from \cite{Nature} is within the
uncertainty of our calculations, which can be understood
qualitatively.

Let us simplify the equation (\ref{eq:extract3}) for the correction.
The first factor in the first term there deviates from unity only by a
few percent and has been neglected in this section. We also explicitly
split the original integral $I_E^0$ into three parts, corresponding
to $I_{3<}^{\rm E}, I_{3>}^{\rm E}-I_3^{\rm R}$ and $I_3^{\rm R}$.
The modified equation for the correction takes the form
\begin{eqnarray}\label{eq:extract3prime}
&&(R_E)^2-(R_E^0)^2 \simeq\frac{24(Z\alpha)m_r}{\pi}\nonumber\\
&\times&\left\{
\int_0^{q_0}{\frac{dq}{q^4}}
\left[\left(\left(G_E(q^2)\right)^2-1-2G_E^\prime(0)\,q^2\right)\right.
\right.\nonumber\\
&&\left.-\left(\left(G_0(q^2)\right)^2-1-2G_0^\prime(0)\,q^2\right)\right]
\nonumber\\
&&+ \int_{q_0}^\infty {\frac{dq}{q^4}}\left[\left(G_E(q^2)\right)^2-\left(G_0(q^2)\right)^2\right]\nonumber\\
&&+\left. \int_{q_0}^\infty {\frac{dq}{q^2}}\frac{R_E^2-(\hat{R}_E^0)^2}{3}\right\}\;,
\end{eqnarray}
where $G_0(q^2)$ is the form factor applied in the former evaluation and
\[
(\hat{R}_E^0)^2=-6 G^\prime_0(0)
\]
is the radius, which follows from $G_0(q^2)$ and which does not
necessarily coincide with $R_E^0$.

The contribution from the low momenta is consistent with zero for both
evaluations (in \cite{Nature} and \cite{Science}) since the integral
\begin{eqnarray}
&&
\int_0^{q_0}{\frac{dq}{q^4}}
\left[\left(\left(G_E(q^2)\right)^2-1-2G_E^\prime(0)\,q^2\right)\right.
\nonumber\\
&&\left.-\left(\left(G_0(q^2)\right)^2-1-2G_0^\prime(0)\,q^2\right)\right]
\nonumber\\
&\simeq& \Bigl(\bigl(1\pm1\bigl)C_{\rm dip}-C_0\Bigr)\,q_0\nonumber
\end{eqnarray}
agrees with zero within the uncertainty we applied. Here $C_0$ is the
coefficient at $q^4$ for $G_0$.

As for the data part
\[
\int_{q_0}^\infty {\frac{dq}{q^4}}\left[\left(G_E(q^2)\right)^2-\left(G_0(q^2)\right)^2\right]\;,
\]
the situation in two former evaluations is different.
In the case of the evaluation applied in \cite{Science}, the fit used
there is consistent with ours. The contribution of this term is
consistent with zero and a departure from zero is below the
uncertainty of our calculation.

\begin{figure}[thbp]
\begin{center}
\resizebox{0.85\columnwidth}{!}{\includegraphics{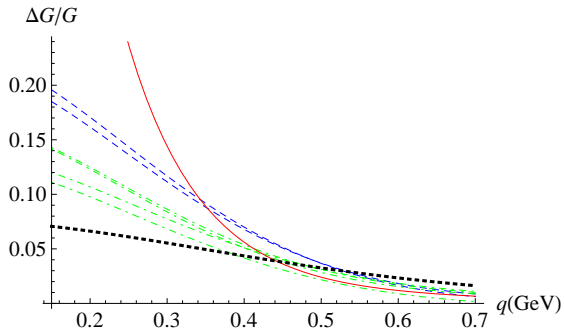}}
\end{center}
\caption{Relative deviation of $(G_E^2-1)/q^4$, the integrand for
$I_{3>}^{\rm E}$, for the dipole fit with $\Lambda$, related to
$R_E=0.8418\;$fm \cite{Nature} (a bold black dotted line),
from the standard dipole fit
with $\Lambda^2=0.71\:{\rm GeV}^2$. Effectively, the former fit was
used in extraction in \cite{Nature}, while the latter dipole fit is
better consistent with the data at high $q$. All the other lines are
for the integrand from the fits used above and we use the same legend
as in the previous plots.}
\label{fig:g2-1}       
\end{figure}

Considering the evaluation of \cite{Nature} {\em a
posteriori\/}, the fit, which was used in there, is the dipole
parametrization with the parameter related to the final value of the
charge radius. (Technically, they used the dipole fit with an
adjustable parameter, but the parameter was determined by the
extracted value of the radius.) For the contribution under question,
a comparison of the
standard dipole fit, which within a few percent is consistent with
the fits we applied here, is plotted in Fig.~\ref{fig:g2-1}. The
discrepancy of the integrand for momenta above $q_0\simeq
0.18\,\Lambda=0.15\,{\rm GeV}/c$ is not more than 10\%, which is
the uncertainty of our calculation of $I_{3>}^{\rm E}-I^{\rm R}_3$. Thus the
results for the correction should be still  consistent with zero.

The last term in (\ref{eq:extract3prime}) is
\[
\int_{q_0}^\infty {\frac{dq}{q^2}}\frac{R_E^2-(\hat{R}_E^0)^2}{3}=\frac{R_E^2-(\hat{R}_E^0)^2}{3q_0}\;.
\]
Since the evaluation in \cite{Nature} is organized in such a way
that the dipole fit has an adjustable parameter, we find
\[
\hat{R}_E^0={R}_E^0\simeq R_E\;.
\]
In other words, neglecting a small correction to former radius, we see
that the radius-related term for the shift is nearly vanishing.
As a result, since two previous contributions are consistent with
zero,
the eventual shift of the re-evaluation of the result from
\cite{Nature}
produces a shift in (\ref{eq:newre}) which is within the uncertainty.

On the contrary, the radius-related term for the re-evaluation
of the result from \cite{Science} does not vanish.
It is not that small. While the radius extracted there is about the
same as that extracted in \cite{Nature}, the fit applied for the
evaluation of $I_3^{\rm E}$ assumed a value of $G^\prime_E(0)$,
which is not consistent with the radius. In other words,
$\hat{R}_E^0$ was quite different there from ${R}_E^0$ of
\cite{Science}.
This produces a non-negligible contribution to the shift. Eventually,
it happens that the correction (\ref{eq:extract3prime}) for the shift
from the result from \cite{Science} somewhat exceeds the uncertainty
of our calculations.

It might look paradoxal and confusing that the
correction for an evaluation with a rather unrealistic fit
for the proton form factor is
substantially smaller than the correction for an evaluation based on a
`good' fit of the data. However, one can see now that the effect
was caused by the inconsistency of  the applied fit
(and the related charge radius from scattering \cite{am2007})
and the value of the charge radius extracted from muonic hydrogen.

Roughly speaking there are two scenarios (without
suggesting `new physics') to explain the discrepancy between the
results from the muonic-hydrogen Lamb shift and the electron-proton scattering.
\begin{itemize}
\item[A).] If the radius from the scattering is correct, that means
that, since theory of muonic hydrogen is well established at the
level of the controversy, the muonic-hydrogen experiment should be
strongly incorrect. In this case there is not much sense to evaluate
the $I_3^{\rm E}$ accurately.
\item[B).] If the experiment on muonic hydrogen is correct (at
the level of the discrepancy), then we have to conclude that the
scattering result for the radius is wrong. The data are more or less
correct. However, statistically evaluating a big set of the data,
one may easily miss systematic effects which being negligible for
each of data points, are important for their set as a whole. The
accuracy of the fit in such a scenario should be overestimated.

One has to clearly distinguish between two kinds of fitting. A
theoretically motivated fit with its shape known {\em a priori\/} is
a way for the best determination of certain physical parameters
and it allows afterwards any operations on the fit. On the contrary, a
phenomenological fit does not determine any physical parameter,
because the fit parameters may have no physical meaning at all. Such a fit
is simply a function which is consistent with the data. Applying such a fit
for interpolations, extrapolations and differentiations is, in
general, questionable. As an example, we refer to the fits applied in
our paper. The difference between the fit (\ref{fit:bo1994}) (with
polynomials in $q$) from the other fits is within a few percent (as seen
from Fig.~\ref{fig:edip}), however behavior at low momentum is
completely wrong and leads to an infinite value of the charge
radius. Actually, considering a small term, linear in $q$, in any
more reasonable fit would not break its consistency as far as such a
term is small. But it should produce absolutely incorrect results
for the charge radius and $I_3^{\rm E}$. Indeed, we should exclude
such a fit because it does not have an analytic behavior at low
$q^2$.
So, the result strongly depends on theoretical constraints
accompanying the fitting procedure. However, there is no reason
to expect that just limiting considerations to
analytic functions at low $q^2$ is sufficient to obtain a
correct extrapolation at low momentum.
There may be other important constraints which may affect the final
results. As we mentioned in the introduction, there is no fit
which is literally correct and is consistent with all the data.

The missed systematic effects can be in the raw data, in their
theoretical evaluation (by correcting for QED effects and proton
polarizability) or just in fitting with an arbitrary class of
functions, which by the way does not reproduce\footnote{One should
not overestimate this issue. The hadronic vacuum polarization in the
two-pion channel can be considered in different ways. A well-known
`realistic' form factor of the pion by Gounaris and Sakurai \cite{GS}
allows to take into account the correct position of the branch point
and the cut line for two-pion production, however, the eventual
result for the vacuum polarization operator at space-like momentum
is not much different from a bold picture
with just a narrow $\rho$-meson pole.} correct analytic behavior at
negative $q^2$.

In this scenario, while the data are roughly correct, the very shape
of the fit is much more uncertain than expected. The value of
$G^\prime_E(0)$ should be linked to the charge radius whatever it is. In
this scenario, such a radius is to be from muonic hydrogen.
\end{itemize}

That means that the difference in results for the complete integral
$I_3^{\rm E}$ between different evaluations (\ref{eq:extract3prime})
is basically due to appropriate
(or non-appropriate) choice of $G_E^\prime(0)$.
The unrealistic fit (the
dipole parametrization with the parameter to be consistent with the
final value of the charge radius as in \cite{Nature}) for the
subtracted form factor has
reasonable behavior at relatively large $q^2$, while the realistic
fit for the subtracted form factor, which produces a higher value of
$R_E$, has  `bad' behavior at
high $q^2$. The `direct' data contribution is relatively unimportant
in this area as seen in Fig~\ref{fig:ind3}.

That explains, why the correction is bigger for the re-evaluation of the more
recent CREMA's result from \cite{Science}, than for the re-evaluation of
the result from \cite{Nature}.

Concluding the discussion on the central value of the correction,
we have to stress that the uncertainty of the correction does not
depend too much on the choice of the model to evaluate the
higher-order proton-finite-size contribution. We strongly
believe that the uncertainty of this term was underestimated in the past.

\section{Conclusions}

Above, we have re-evaluated the contribution of $I_3^{\rm E}$ only. All
other details have not been reconsidered.
The final value of the
proton charge radius should be obtained from the re-evaluation of the
more recent CREMA publication \cite{Science}, which has certain advantages
from the experimental and theoretical point of view.

The new value of the radius obtained here is
\[
R_E = 0.840\,22(56)\;{\rm fm}\,.
\]
The uncertainty is bigger than in the original publication
\cite{Science}.

\begin{figure}[thbp]
\begin{center}
\resizebox{0.90\columnwidth}{!}{\includegraphics{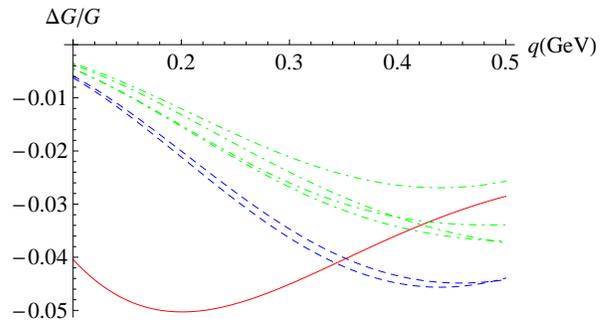}}
\end{center}
\caption{Relative deviation of the electric form factor from the
dipole form factor $(G_E-G_{\rm dip})/G_{\rm dip}$. The horizontal axis:
$q\;$[Gev/$c$].
}
\label{fig:40}       
\end{figure}

The scatter in the calculations of $I_>$ is comparable with our
estimation for the uncertainty and it is rather  smaller than our
estimation of $\delta I_{3>}^{\rm E}$ (if we exclude
(\ref{fit:bo1994})). Including
the fit (\ref{fit:bo1994}) with inappropriate behavior at low $q$, we
increase the scatter, but not too much. Behavior of various fits in
the region suitable for the choice of the separation parameter $q_0$ is
presented in Fig.~\ref{fig:40}.

Our estimation of the uncertainty is arbitrary to a certain extent,
but seems reasonable. The estimation of the uncertainty of the form
factor at $q_0$ as 1\% is validated by the behavior of the fits and
the obtained scatter of the results. Nevertheless, a direct
examination of the experimental data around potential values of
$q_0$, a verification of their uncertainty and, if possible, its
improvement below the level of 1\% would be appreciated.

The same could be concluded about an estimation of the $q^4$ term as
within a possible 100\% deviation from the dipole value.
It is validated by the
scatter of the results of various fits applied here. It is consistent with fits considered by MAMI \cite{mami} (see \cite{thesis} for details.) as well. Such an estimation also has sense if one
recalls a geometrical meaning of the form factor at low $q^2$. Due to
that one may expect that its behavior is basically determined by a
single parameter, namely the characteristic size of the proton,
which is a quite compact object.

The situation may look somewhat similar to that for the
anomalous magnetic moment of muon. Any precision QED calculations
are incomplete and at certain stage one has to deal with hadronic
effects.
In particular, in case of $g-2$ of the muon and the hyperfine interval
in muonium the hadronic contributions are presented with the hadronic
vacuum polarization. 
One may think that dealing with the proton form factor
while calculating the energy levels in muonic hydrogen is similar to
the treatment of the hadronic vacuum polarization effects.
However, this impression is incorrect.
Integration for the hadronic vacuum polarization is done
\cite{amu:book2,prd}
directly over the data with no subtractions or extrapolations.
That makes such evaluations reliable. The involvement of the
subtractions and extrapolations in case of the
higher-order proton-finite-size corrections has changed the
situation completely and requires a different considerations,
which is presented in this paper.

The author is grateful to S. Eidelman and V. Ivanov
for useful discussions. This work was supported in part by
DFG under grant HA 1457/9-1.

\appendix

\section{Fits for the electric form factor of the proton
applied in the paper\label{app:fit}\label{s:fit}}

The fits for $G_E$ applied in the paper include two chain-fraction
fits.
Both are from Arrington and Sick, 2007, \cite{as2007}.
The first one is completely based on \cite{as2007}%
\footnote{Here, $Q$ is the numerical
value for the momentum transfer $q$ in GeV.}
\begin{eqnarray}\label{fit:as2007}
G_E(q^2)&=& \frac{1}{1 + \frac{3.44 Q^2}{1 - \frac{0.178 Q^2}{ 1 -
\frac{1.212 Q^2}{1 +
\frac{1.176Q^2}{1 - 0.284Q^2 } } } } }\;,
\end{eqnarray}
while the other is obtained in \cite{as2007} by applying
the two-photon correction according to \cite{bm2005}
\begin{eqnarray}\label{fit:bm2005}
G_E(q^2)&=& \frac{1}{1 + \frac{3.478Q^2}{
        1 - \frac{0.140 Q^2 }{1 - \frac{1.311Q^2}{1 + \frac{1.128Q^2}{1 - 0.233Q^2}}}}}\;.
\end{eqnarray}


Four fits are with the Pad\'e approximation in $q^2$,
originally introduced by Kelly, 2004,
\cite{kelly}
\begin{eqnarray}\label{fit:kelly}
G_E&=&\frac{1 - 0.24 \tau }{1 + 10.98 \tau + 12.82 \tau^2 +
      0.863 \tau^3}\;,
\end{eqnarray}
where
\[
\tau = q^2/4m_p^2\;,
\]
and later developed by Arrington 
et al., 2007, \cite{am2007}
\begin{eqnarray}\label{fit:am2007}
G_E&=&\frac{1 + 3.439 \tau - 1.602\tau^2 + 0.068 \tau^3}{D_A}\;,\nonumber\\
D_A&=&1 + 15.055 \tau + 48.061 \tau^2 +
      99.304 \tau^3\nonumber\\
&& + 0.012  \tau^4 + 8.650\tau ^5\;,
\end{eqnarray}
Alberico 
et al., 2009, \cite{ab2009}
\begin{eqnarray}\label{fit:ab2009}
G_E(q^2)&=& \frac{1 - 0.19\tau}{1 + 11.12 \tau + 15.16\tau^2 + 21.25
\tau^3}\;,
\end{eqnarray}
and Venkat 
et al., 2011, \cite{va2011}
\begin{eqnarray}\label{fit:va2011}
G_E&=&\frac{N_V}{D_V}\;,\nonumber\\
N_V&=&1+2.909\,66\tau-1.115\,422\,29\tau^2+3.866\,171\times10^{-2}\tau^3\nonumber\\
D_V&=&1+14.518\,7212\tau+40.883\,33\tau^2+99.999\,998\tau^3\nonumber\\
&& +4.579\times10^{-5}\tau^4+10.358\,0447\tau^5\;.
\end{eqnarray}

One more fit with is the Pad\'e approximation in $q$ from Bosted, 1995,
\cite{bo1994}
\begin{eqnarray}\label{fit:bo1994}
G_E(q^2)&=& \frac{1}{1 + 0.62Q + 0.68 Q^2 + 2.8 Q^3 + 0.83Q^4}\;.~~~~~~
\end{eqnarray}

\section{Evaluation of $I_3^{\rm E}$ in former extractions
of $R_E$ from the Lamb shift in muonic hydrogen}

\subsection{Evaluation in \cite{Nature}\label{s:nat}}

The calculation of $I_3^{\rm E}$ in \cite{Nature} was done by applying
the dipole parametrization with a free parameter and allowing this
parameter to be consistent with the radius extracted (see, e.g.,
\cite{EGS}).

\subsection{Evaluation in \cite{Science}\label{s:sci}}

The result for the evaluation of $I_3^{\rm E}$, applied
in \cite{Science}, was not presented there directly,
being a part of the adopted there value for the two-photon-exchange
correction. The result for the two-photon-exchange was
taken from \cite{birse}. That result is a sum of `elastic' and
polarizability contributions, with the former strongly dominating.
Meantime, the `elastic' term was not calculated there, but taken
from \cite{carlson}. Their calculation of the `elastic' term takes
into account recoil effects. Nevertheless, the leading contribution
is a non-recoil one, which is determined by $I_3^{\rm E}$. The value
of integral was found by integrating over the fit (\ref{fit:am2007})
from \cite{am2007}. The uncertainty was estimated via a comparison
with results from some other fits, such as (\ref{fit:kelly})
\cite{kelly}.

Thus, the leading model-dependent effect due to the form factors is
still from the calculation of $I_3^{\rm E}$, the central value of
which was obtained by applying (\ref{fit:am2007}).


\begin{thebibliography}{00.}

\frenchspacing

\bibitem{Nature}
R. Pohl, A. Antognini, F. Nez, F.D. Amaro, F. Biraben, J.M.R.
Cardoso, D.S. Covita, A. Dax, S. Dhawan, L.M.P. Fernandes, A.
Giesen, T. Graf, T.W. H\"ansch, P. Indelicato, L. Julien, Cheng-Yang
Kao, P. Knowles, E.-O. Le Bigot, Yi-Wei Liu, J.A.M. Lopes, L.
Ludhova, C.M.B. Monteiro, F. Mulhauser, T. Nebel, P. Rabinowitz,
J.M.F. dos Santos, L.A. Schaller, K. Schuhmann, C. Schwob, D. Taqqu,
J.F.C.A. Veloso and F. Kottmann, Nature (London) {\bf 466}, 213
(2010).

\bibitem{Science} A.~Antognini, F. Nez, K. Schuhmann, F.D. Amaro, F. Biraben,
J.M.R. Cardoso, D.S. Covita, A. Dax, S. Dhawan, M. Diepold, L.M.P.
Fernandes, A. Giesen, A.L. Gouvea, T. Graf, T.W. H\"ansch, P.
Indelicato, L. Julien, Cheng-Yang Kao, P. Knowles, F. Kottmann,
E.-O. Le Bigot, Yi-Wei Liu, J.A.M. Lopes, L. Ludhova, C.M.B.
Monteiro, F. Mulhauser, T. Nebel, P. Rabinowitz, J.M.F. dos Santos,
L.A. Schaller, C. Schwob, D. Taqqu, J.F.C. A. Veloso, J. Vogelsang,
R. Pohl, Science, {\bf 339} 417 (2013).

\bibitem{sick} I. Sick, Phys. Lett. B {\bf 576}, 62 (2003). 


\bibitem{mami} J.C. Bernauer, P. Achenbach, C. Ayerbe Gayoso, R. B\"ohm, D.
Bosnar, L. Debenjak, M.O. Distler, L. Doria, A. Esser, H.
Fonvieille, J.M. Friedrich, J. Friedrich, M. G\'omez Rodr\'iguez de
la Paz, M. Makek, H. Merkel, D.G. Middleton, U. M\"uller, L.
Nungesser, J. Pochodzalla, M. Potokar, S. S\'anchez Majos, B.S.
Schlimme, S. \v{S}irca, Th. Walcher, and M. Weinriefer,  Phys. Rev.
Lett. {\bf 105}, 242001 (2010).

\bibitem{codata2010}
P.J. Mohr, B.N. Taylor, and D.B. Newell,
Rev. Mod. Phys. {\bf 84}, 1527 (2012).


\bibitem{my_adp} S.G.~Karshenboim, Annalen der Physik {\bf 525}, 472 (2013).

\bibitem{my_ufn} S.G.~Karshenboim, Physics-Uspekhi {\bf 56}, 883 (2013).

\bibitem{EGS} M.I. Eides, H. Grotch, and V.A. Shelyuto, {\em Theory of
Light Hydrogenic Bound States\/}, Springer Tracts Mod. Phys. {\bf
222} (Springer, Berlin, Heidelberg, 2007).


\bibitem{kelly} J.J. Kelly, Phys. Rev. C {\bf 70}, 068202 (2004).


\bibitem{as2007} J. Arrington and I. Sick, Phys. Rev. C{\bf 76}, 035201
(2007).


\bibitem{am2007} J. Arrington, W. Melnitchouk, and J.A. Tjon,  Phys. Rev. C{\bf 76}, 035205
(2007).

\bibitem{ab2009} W.M. Alberico, S.M. Bilenky, C. Guinti, and K.M. Graczyk,  Phys. Rev. C{\bf 79},
065204 (2009).


\bibitem{va2011} S. Venkat, J. Arrington, G.A. Miller and X. Zhan,
Phys. Rev. C{\bf 83}, 015203 (2011).

\bibitem{lorenz} I.T. Lorenz, H.W. Hammer, U.G. Meissner, Eur. Phys. J.  A{\bf 48}, 151
(2012).


\bibitem{pla} S.G.~Karshenboim, Phys. Lett. A{\bf 225}, 97 (1997).


\bibitem{bo1994} P.E. Bosted, Phys. Rev. C{\bf 51}, 409 (1995).




\bibitem{carlson} C.E. Carlson and M. Vanderhaeghen, Phys. Rev. A{\bf 84},
020102 (2011).


\bibitem{birse} M.C. Birse and J.A. McGovern, Eur. Phys. J. A{\bf 48}, 120 (2012).



\bibitem{GS} G.I.~Gounaris and J.J.~Sakurai, Phys. Rev. Lett. {\bf 21} 244
(1968).


\bibitem{thesis} J. Bernauer, {\em Measurement of the elastic electron-proton cross section and separation of the electric and magnetic form factor in the $Q^2$ range from 0.004 to 1\;$({\rm GeV}/c)^2$.\/} Ph.D. Thesis, Mainz, 2010. Available at http://wwwa1.kph.uni-mainz.de/A1/publications/doctor/bernauer.pdf.

\bibitem{amu:book2} F. Jegerlehner, {\em The Anomalous Magnetic Moment of the
Muon\/}. Springer, Berlin and Heidelberg: Springer Tracts in Modern
Physics {\bf 226} (2007).


\bibitem{prd} A. Czarnecki, S. I. Eidelman and S. G. Karshenboim, Phys. Rev. D{\bf 65} (2002).


\bibitem{bm2005} P.G. Blunden, W. Melnitchouk, and J.A. Tjon,  Phys. Rev. C{\bf 72},
034612 (2005).

\end{thebibliography}
\end{document}